\documentclass[10pt]{blois07}
\usepackage{graphicx}
\usepackage{cite,./mcite}
\usepackage{axodraw}
\usepackage{pstricks}
\usepackage{color}
\newcommand\naive{na\"{\i}ve}
\newcommand\as{\alpha_s}
\begin{document}
\begin{flushright}
CERN--PH--TH/2007--176\\
MAN/HEP/2007/22
\end{flushright}
\begin{center}
\vspace*{5cm}
{\Large\bf Breakdown of Coherence ?}
\vspace*{1cm}\\
{\large\bf Michael H. Seymour}
\vspace*{0.5cm}\\
{Physics Department, CERN, CH-1211 Geneva 23, Switzerland,\\ 
and School of Physics \& Astronomy, University of Manchester, Manchester,
M13 9PL, U.K.}
\vspace*{1cm}\\
\begin{quote}
In a recent paper, Albrecht Kyrieleis, Jeff Forshaw
and I discovered a new tower of super-leading logarithms in
gaps between jets cross sections.  After discussions with the referee of
our paper and further investigation, we have come to view this as a
breakdown of naive coherence for initial state radiation.  In this
talk I illustrate this statement in a simple way, and show how it
results in the super-leading logarithms.
\end{quote}
\vspace*{\fill}
{\it Talk given at 12th International Conference on Elastic and
Diffractive Scattering (EDS '07), May 21st--25th 2007, DESY, Hamburg.}
\end{center}
\newpage\mbox{}\newpage
\title{Breakdown of Coherence ?}
\author{Michael H. Seymour$^{1,2}$\protect\footnote{~~~talk presented at EDS07}}
\institute{$^1$Physics Department, CERN, CH-1211 Geneva 23, Switzerland,\\ 
$^2$School of Physics \& Astronomy, University of Manchester, Manchester,
M13 9PL, U.K.}
\maketitle
\begin{abstract}
In a recent paper\cite{Forshaw:2006fk}, Albrecht Kyrieleis, Jeff Forshaw
and I discovered a new tower of \emph{super-leading\/} logarithms in
gaps between jets cross sections.  After discussions with the referee of
our paper and further investigation, we have come to view this as a
breakdown of \naive\ coherence for initial state radiation.  In this
talk I illustrate this statement in a simple way, and show how it
results in the super-leading logarithms.
\end{abstract}

\section{Introduction and The Bottom Line}
\begin{figure}[t]
\centerline{%
\SetScale{0.25}\begin{picture}(190,90) (-30,-30)
    \put(-27,-30){\scalebox{0.5}{\includegraphics{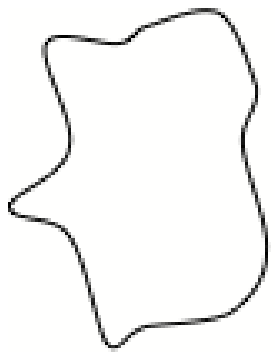}}}
    \put(-30,30){(a)}
    \SetWidth{4.0}
    \SetColor{Black}
    \ArrowLine(30.14,-3.62)(391.76,104.87)
    \ArrowLine(391.76,128.98)(632.84,225.41)
    \ArrowLine(391.76,116.92)(644.9,201.3)
    \ArrowLine(391.76,104.87)(644.9,177.2)
    \ArrowLine(391.76,92.82)(644.9,153.09)
    \ArrowLine(391.76,80.76)(644.9,128.98)
    \GOval(391.76,104.87)(60.27,24.11)(0){0.882}
  \end{picture}
}\vspace*{1cm}
\centerline{%
\SetScale{0.25}\begin{picture}(190,90) (-30,-30)
    \put(-27,-30){\scalebox{0.5}{\includegraphics{blob.ps}}}
    \put(-30,30){(b)}
    \put(-45,-10){$\displaystyle\sum_i$}
    \put(165,50){$i$}
    \SetWidth{4.0}
    \SetColor{Black}
    \ArrowLine(30.14,-3.62)(391.76,104.87)
    \ArrowLine(391.76,128.98)(632.84,225.41)
    \ArrowLine(391.76,116.92)(644.9,201.3)
    \ArrowLine(391.76,104.87)(644.9,177.2)
    \ArrowLine(391.76,92.82)(644.9,153.09)
    \ArrowLine(391.76,80.76)(644.9,128.98)
    \GOval(391.76,104.87)(60.27,24.11)(0){0.882}
    \Gluon(554.49,168.15)(590.65,-0.6){18.08}{4}
    \SetWidth{0.5}
    \Vertex(554.49,168.15){8.63}
  \end{picture}
\hfill
\SetScale{0.25}\begin{picture}(190,90) (-30,-6.65)
    \put(-27,-6.65){\scalebox{0.5}{\includegraphics{blob.ps}}}
    \put(-30,53.35){(c)}
    \put(-50,35){$\equiv$}
    \SetWidth{4.0}
    \SetColor{Black}
    \ArrowLine(30.14,89.8)(391.76,198.29)
    \ArrowLine(391.76,222.4)(632.84,318.83)
    \ArrowLine(391.76,210.34)(644.9,294.72)
    \ArrowLine(391.76,198.29)(644.9,270.61)
    \ArrowLine(391.76,186.24)(644.9,246.51)
    \ArrowLine(391.76,174.18)(644.9,222.4)
    \GOval(391.76,198.29)(60.27,24.11)(0){0.882}
    \Gluon(277.24,165.14)(313.41,-3.62){18.08}{4}
    \SetWidth{0.5}
    \Vertex(277.24,165.14){8.63}
  \end{picture}
}
\caption{Illustration of \naive\ coherence in final-state radiation.
A hard parton produced in the hard process fragments into a system of
hard collinear partons (a).  The amplitude for this system to emit a
soft, wide-angle, real or virtual gluon should be calculated from the
insertion of the soft gluon onto each of the external hard partons,
summed over these partons (b).  Colour coherence implies that this can
be calculated \emph{as if\/} the soft gluon were emitted by the
original hard parton, i.e.~by the total colour charge of the jet~(c).}
\label{fig1}
\end{figure}

I begin by illustrating, in a simple pictorial way, what I mean by
\naive\ coherence.  Consider an arbitrary hard process that produces a
hard parton, which then fragments into a system of hard collinear
partons, as shown in Fig.~\ref{fig1}a.  To be precise, by hard collinear
I mean that the plus components of all the partons are of the same order
as that of the originating parton, and all their transverse momenta are
much smaller, with the originating parton defining the plus direction.
Consider calculating the first correction to this amplitude coming from
a soft wide-angle gluon.  Again, to be precise, by soft wide-angle, I
mean that its transverse momentum is much smaller than the relative
transverse momenta of all collinear partons in the jet, and that its
plus momentum is at most of order its transverse momentum.  As
illustrated in Fig.~\ref{fig1}b, this amplitude is obtained from the
first one by inserting the soft wide-angle gluon onto each of the
external partons, summing over those partons.  Studying the integral
over the momentum of the soft wide-angle gluon, it is straightforward to
see that the momentum-dependent parts of all these insertions are
identical and they only differ by colour algebra.  It is also
straightforward to show, for example using the diagrammatic technique of
\cite{Dokshitzer:1985vs,*Hakkinen:1996bb}, that the contributions are
simply additive in colour space.  The final result is therefore, as
illustrated in Fig.~\ref{fig1}c, that the amplitude can be calculated
\emph{as if\/} the soft wide angle gluon was emitted by an
\emph{on-shell\/} parton with the same plus momentum and colour as the
initiating parton.  This is the usual statement of \naive\ colour
coherence: soft wide angle gluons are emitted by the jet as a whole,
imagined to be on shell.

\begin{figure}[tp]
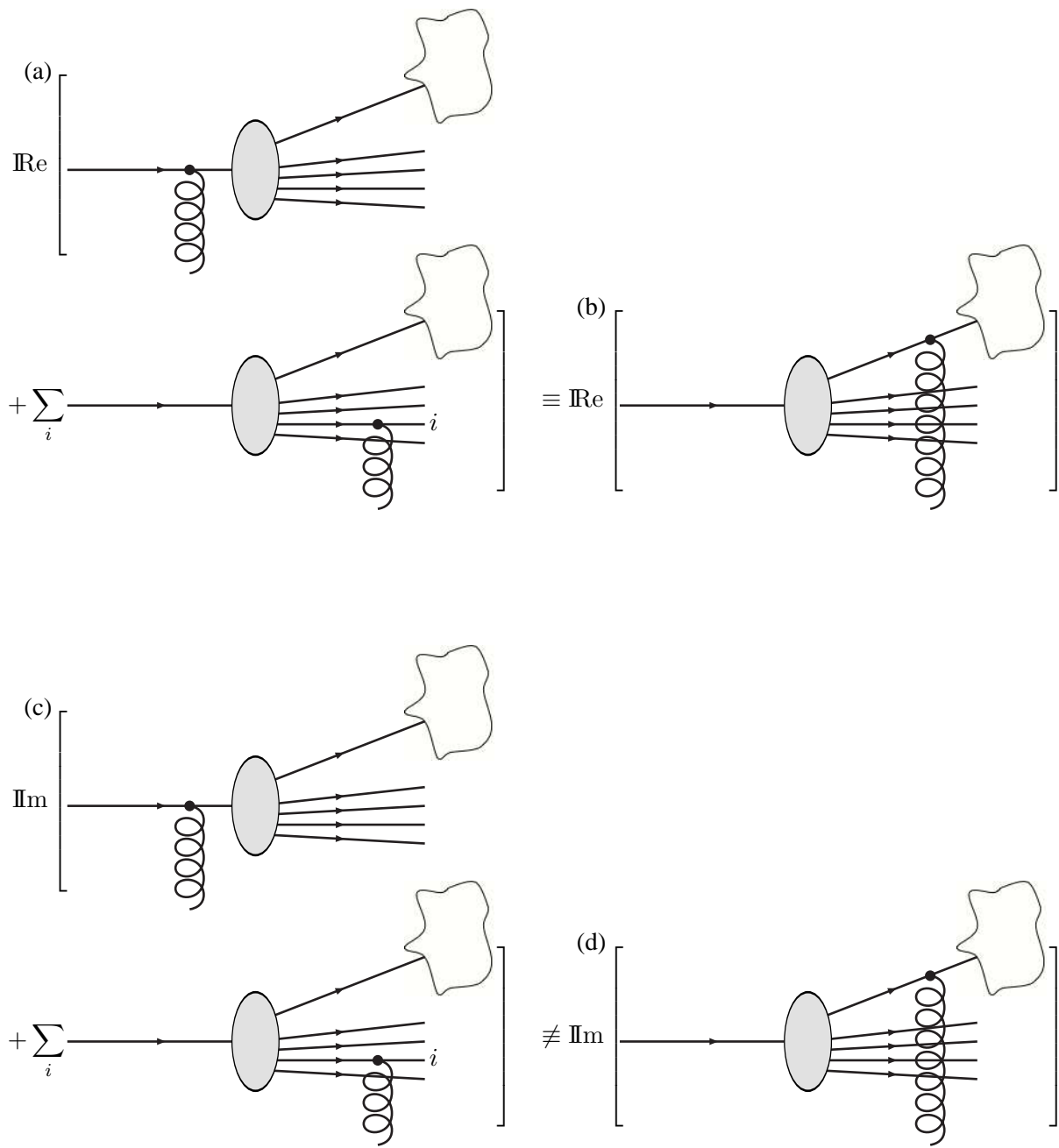

\centerline{%
\SetScale{0.25}\begin{picture}(180,115) (48,-4)
    \put(194,60){\scalebox{0.5}{\includegraphics{blob.ps}}}
    \put(30,80){(a)}
    \put(25,40){$\mathrm{I\!Re}\left[\phantom{\raisebox{1.5cm}{.}}\right.$}
    \SetWidth{4.0}
    \SetColor{Black}
    \ArrowLine(194.13,161.78)(517.69,161.78)
    \ArrowLine(517.69,194.13)(808.89,307.38)
    \ArrowLine(517.69,113.25)(808.89,97.07)
    \ArrowLine(517.69,129.42)(808.89,129.42)
    \ArrowLine(517.69,145.6)(808.89,161.78)
    \ArrowLine(517.69,161.78)(808.89,194.13)
    \GOval(517.69,161.78)(84.93,40.44)(0){0.882}
    \Gluon(404.45,161.78)(404.45,-16.18){24.27}{4}
    \SetWidth{0.5}
    \Vertex(404.45,161.78){9.15}
  \end{picture}
\hfill
\SetScale{0.25}\begin{picture}(180,115) (48,-4)
  \end{picture}
}\vspace*{-0.5cm}
\centerline{%
\SetScale{0.25}\begin{picture}(180,115) (48,-4)
    \put(194,60){\scalebox{0.5}{\includegraphics{blob.ps}}}
    \put(23,38){$+\displaystyle\sum_i$}
    \put(205,30){$i$}
    \put(230,40){$\left.\phantom{\raisebox{1.5cm}{.}}\right]$}
    \SetWidth{4.0}
    \SetColor{Black}
    \ArrowLine(194.13,161.78)(517.69,161.78)
    \ArrowLine(517.69,194.13)(808.89,307.38)
    \ArrowLine(517.69,113.25)(808.89,97.07)
    \ArrowLine(517.69,129.42)(808.89,129.42)
    \ArrowLine(517.69,145.6)(808.89,161.78)
    \ArrowLine(517.69,161.78)(808.89,194.13)
    \GOval(517.69,161.78)(84.93,40.44)(0){0.882}
    \Gluon(728.01,129.42)(728.01,-16.18){24.27}{3}
    \SetWidth{0.5}
    \Vertex(728.01,129.42){9.15}
  \end{picture}
\hfill
\SetScale{0.25}\begin{picture}(180,115) (48,-4)
    \put(194,60){\scalebox{0.5}{\includegraphics{blob.ps}}}
    \put(30,80){(b)}
    \put(15,40){$\equiv\mathrm{I\!Re}\left[\phantom{\raisebox{1.5cm}{.}}\right.$}
    \put(230,40){$\left.\phantom{\raisebox{1.5cm}{.}}\right]$}
    \SetWidth{4.0}
    \SetColor{Black}
    \ArrowLine(194.13,161.78)(517.69,161.78)
    \ArrowLine(517.69,194.13)(808.89,307.38)
    \ArrowLine(517.69,113.25)(808.89,97.07)
    \ArrowLine(517.69,129.42)(808.89,129.42)
    \ArrowLine(517.69,145.6)(808.89,161.78)
    \ArrowLine(517.69,161.78)(808.89,194.13)
    \GOval(517.69,161.78)(84.93,40.44)(0){0.882}
    \Gluon(728.01,275.02)(728.01,-16.18){24.27}{7}
    \SetWidth{0.5}
    \Vertex(728.01,275.02){9.15}
  \end{picture}
}\vspace*{2cm}
\centerline{%
\SetScale{0.25}\begin{picture}(180,115) (48,-4)
    \put(194,60){\scalebox{0.5}{\includegraphics{blob.ps}}}
    \put(30,80){(c)}
    \put(25,40){$\mathrm{I\!Im}\left[\phantom{\raisebox{1.5cm}{.}}\right.$}
    \SetWidth{4.0}
    \SetColor{Black}
    \ArrowLine(194.13,161.78)(517.69,161.78)
    \ArrowLine(517.69,194.13)(808.89,307.38)
    \ArrowLine(517.69,113.25)(808.89,97.07)
    \ArrowLine(517.69,129.42)(808.89,129.42)
    \ArrowLine(517.69,145.6)(808.89,161.78)
    \ArrowLine(517.69,161.78)(808.89,194.13)
    \GOval(517.69,161.78)(84.93,40.44)(0){0.882}
    \Gluon(404.45,161.78)(404.45,-16.18){24.27}{4}
    \SetWidth{0.5}
    \Vertex(404.45,161.78){9.15}
  \end{picture}
\hfill
\SetScale{0.25}\begin{picture}(180,115) (48,-4)
  \end{picture}
}\vspace*{-0.5cm}
\centerline{%
\SetScale{0.25}\begin{picture}(180,115) (48,-4)
    \put(194,60){\scalebox{0.5}{\includegraphics{blob.ps}}}
    \put(23,38){$+\displaystyle\sum_i$}
    \put(205,30){$i$}
    \put(230,40){$\left.\phantom{\raisebox{1.5cm}{.}}\right]$}
    \SetWidth{4.0}
    \SetColor{Black}
    \ArrowLine(194.13,161.78)(517.69,161.78)
    \ArrowLine(517.69,194.13)(808.89,307.38)
    \ArrowLine(517.69,113.25)(808.89,97.07)
    \ArrowLine(517.69,129.42)(808.89,129.42)
    \ArrowLine(517.69,145.6)(808.89,161.78)
    \ArrowLine(517.69,161.78)(808.89,194.13)
    \GOval(517.69,161.78)(84.93,40.44)(0){0.882}
    \Gluon(728.01,129.42)(728.01,-16.18){24.27}{3}
    \SetWidth{0.5}
    \Vertex(728.01,129.42){9.15}
  \end{picture}
\hfill
\SetScale{0.25}\begin{picture}(180,115) (48,-4)
    \put(194,60){\scalebox{0.5}{\includegraphics{blob.ps}}}
    \put(30,80){(d)}
    \put(15,40){$\not\equiv\mathrm{I\!Im}\left[\phantom{\raisebox{1.5cm}{.}}\right.$}
    \put(230,40){$\left.\phantom{\raisebox{1.5cm}{.}}\right]$}
    \SetWidth{4.0}
    \SetColor{Black}
    \ArrowLine(194.13,161.78)(517.69,161.78)
    \ArrowLine(517.69,194.13)(808.89,307.38)
    \ArrowLine(517.69,113.25)(808.89,97.07)
    \ArrowLine(517.69,129.42)(808.89,129.42)
    \ArrowLine(517.69,145.6)(808.89,161.78)
    \ArrowLine(517.69,161.78)(808.89,194.13)
    \GOval(517.69,161.78)(84.93,40.44)(0){0.882}
    \Gluon(728.01,275.02)(728.01,-16.18){24.27}{7}
    \SetWidth{0.5}
    \Vertex(728.01,275.02){9.15}
  \end{picture}
}
\caption{The analogue of Fig.~\ref{fig1} for initial-state radiation.
A hard initial-state parton entering the hard process fragments into a
system of hard final-state collinear partons.  The amplitude for this
system to emit a soft, wide-angle, real or virtual gluon should again be
calculated from the insertion of the soft gluon onto each of the
external hard partons, summed over these partons (a,c).  Because when
the soft gluon is virtual the imaginary part of the loop correction is
sensitive to the direction of the momentum flow, the colour coherence
argument can only be used for real emission and the real part of the
loop (b) but not for the imaginary part (d).}
\label{fig2}
\end{figure}

Now I turn to the case of an initial-state parton, Fig.~\ref{fig2}.
Consider an arbitrary hard process initiated by a hard parton, which
fragments into a system of hard collinear partons and its correction
coming from a soft wide-angle gluon, as shown in
Figs.~\ref{fig2}a~and~c.  At first sight it looks the same as the
final-state case and, in fact, if the soft wide-angle gluon is real, it
is, so it is as if the soft wide-angle gluon was emitted by the internal
line, imagined to be on shell, Fig.~\ref{fig2}b.  However, if the soft
wide-angle gluon is virtual, one has to consider the momentum structure
of the loop integral more carefully.  Performing one integration by
contour, we generally pick up poles from either the soft gluon
propagator or the hard parton propagators.  The former gives a real part
that has an identical form in all cases.  The problem then reduces to
colour algebra again and, just like for real emission, it is as if the
soft wide-angle gluon was emitted by the internal line, imagined to be
on shell, Fig.~\ref{fig2}b.  However, for the other pole, coming from
hard parton propagators, its causal structure depends on whether the
hard partons the gluon is attached to are in the final state or the
initial state.  In particular, the imaginary part is zero if the gluon
connects an initial-state parton to a final-state parton\footnote{I am
working in Feynman gauge.}, and non-zero for initial-initial and
final-final connections.  Therefore there is a mismatch between the
different diagrams in Fig.~\ref{fig2}c and they do not correspond to the
contribution from a single on-shell parton, Fig.~\ref{fig2}d.  It is
this statement that we describe as a \emph{breakdown of \naive\
coherence for initial-state radiation}.

We `discovered' this breakdown of coherence in calculating corrections
to the conventional calculations of gaps-between-jets cross sections
from one gluon emitted outside the gap accompanied by any number
of soft wide-angle gluons.  It was a great surprise to us, but we soon
learnt that it was actually well known to the early pioneers of QCD.  In
particular, there are lengthy discussions in the literature of whether
or not these known effects (coming from ``Coulomb gluons'') lead to
violations of the Bloch--Nordsiek theorem (see for example
Ref.~\cite{Doria:1980ak}).  These issues were eventually settled, at
least for massless partons, by Collins, Soper and Sterman's proof of
factorization\cite{Collins:1985ue,*Collins:1988ig}.  The hard collinear,
and soft real, corrections are quickly dealt with in their paper, and
most of the subtlety of their proof is related to gluons with plus and
minus momenta much smaller than their transverse momenta (the ``Glauber
region''), which are exactly the ones that give the imaginary parts we
are discussing.  They showed that these do lead to violations of
factorization in individual diagrams, but that, eventually, these
violations cancel each other after summing over all diagrams \emph{for
the scattering of colour-singlet incoming hadrons}.  Diagrams in which
the gluons are attached to the outgoing hadron remnants are essential
for this cancelation.

However, in calculating perturbatively-exclusive cross sections, for
example the gaps-between-jets cross section defined below, one can
perform factorization at the perturbative scale defined by the scale
below which the observable is inclusive, and one can calculate the cross
section perturbatively using incoming partons defined at this scale.
Therefore one cannot appeal to the hadron remnants, and these effects
really remain in the cross section.

\section{Consequences: Super-Leading Logarithms in Gaps Between Jets
Cross Sections}

In the remainder of the talk, I discuss the consequences of this
breakdown of \naive\ coherence and, in particular, the appearance of
super-leading logarithms in the gaps-between-jets cross section.  Here I
am simply recapping the results of Ref.~\cite{Forshaw:2006fk}, so I can
be brief.

To define the gap cross section, and the kinematic variables I use to
describe it, consider two-jet production at lowest order in hadron
collisions.  Since I am interested in the soft or collinear corrections,
the lowest-order kinematics are sufficient.  I define the jets to have
transverse momenta $Q$ and to be separated by a (reasonably large)
rapidity interval $\Delta y$.  I define a `gap' event sample by summing
up the total scalar transverse momentum in a rapidity interval of length
$Y<\Delta y$ in the region between the two jets and only accepting
events in which this summed transverse momentum is less than $Q_0\le Q$.
Provided $Q_0$ is well above the confinement scale, this gap cross
section is perturbatively calculable.  For $Q_0\ll Q$ it develops large
logarithmic corrections at every order that must be summed to all orders
to yield a reliable result.

The conventional wisdom for such calculations is that the logarithmic
series is $\as^n\log^n$, which define the leading logs for this process,
that these leading logs can be calculated by considering only soft
wide-angle virtual gluons stretched between the hard external partons,
and that for every real emission outside the gap there is an equal and
opposite virtual correction.  Our findings contradict all of these
points: we find super-leading\footnote{I should clear up one possible
point of confusion: by super-leading we do not mean that there are more
than the two logarithms per order expected from QCD, but only that this
observable, which is expected to have only soft contributions, so one
logarithm per power of $\as$, actually develops additional collinear
logarithms at high orders, which we call super-leading since they are
beyond the expected soft-only tower.  More precisely, one power of $Y$
that appears in the coefficient of the leading logarithm, and is the
remnant of the collinear logarithm, gets promoted to become a logarithm
of $Q/Q_0$.} logarithms $\as^n\log^{n+1}$.  We already expected, based
on the work of Dasgupta and Salam\cite{Dasgupta:2002bw} contributions
from emission from gluons outside the gap, but in distinction to their
result which is an edge effect: emission just outside the gap produces
radiation just inside, the effect we find comes from emission
arbitrarily far outside the gap.  These results are directly related to
a real--virtual mis-cancellation due to Coulomb gluon effects and
ultimately due to the breakdown of \naive\ coherence for initial-state
radiation.

To illustrate how these effects ultimately give rise to the
super-leading logarithms, I briefly recap the ingredients of the
`conventional' calculations for gaps-between-jets developed by Sterman
and others over many
years\cite{Sotiropoulos:1993rd,*Oderda:1998en,*Berger:2001ns}, first in
the simpler setting of $e^+e^-$ annihilation.

\subsection{Gaps in \boldmath$e^+e^-$ annihilation}

The lowest order process produces a quark of momentum $p_1$ and an
antiquark of momentum $p_2$.  Its amplitude is defined to be
$\mathcal{A}_0$.  The one-loop correction in the Feynman gauge is given by
the single diagram with a gluon of momentum $k$ stretched between $p_1$
and $p_2$.  In order to extract the leading logarithms, the eikonal
approximation is sufficient.  Performing the loop integral over $k$ by
contour, one picks up poles at $k^2=0$ and at $p_1^2=p_2^2=0$.  The
former gives a contribution that has exactly the form of a phase-space
integral for a real gluon emission and leads to a term in the cross
section that is exactly equal and opposite to the real-emission cross
section.  The conventional calculation uses this fact, by assuming that
the real--virtual cancellation is perfect for transverse momenta below
$Q_0$ and rapidities outside the gap, so that the entire first-order
correction can be calculated from the loop diagram integrated over the
disallowed region of phase space,
\begin{equation}
\mathcal{A}_1=-\frac{2\alpha_s}{\pi}
\int_{Q_0}^Q\frac{dk_t}{k_t}C_F
\left(Y-i\pi\right)
\mathcal{A}_0,
\end{equation}
where the $Y$ term is the integral of the $k^2=0$ pole over the gap
region and the $i\pi$ term comes from the $p_1^2=p_2^2=0$ pole.  To
obtain the leading logarithmic contribution at $n$th order, one can
simply nest the $k_t$ integral $n$ times and obtain
\begin{equation}
\mathcal{A}=e^{-\frac{2\alpha_s}{\pi}
\int_{Q_0}^Q\frac{dk_t}{k_t}C_F
\left(Y-i\pi\right)}
\mathcal{A}_0.
\end{equation}
The gap cross section is then given by
\begin{equation}
\sigma=\mathcal{A}^\star\mathcal{A}=
\mathcal{A}_0^\star
e^{-\frac{2\alpha_s}{\pi}
\int_{Q_0}^Q\frac{dk_t}{k_t}C_F
\left(Y+i\pi\right)}
e^{-\frac{2\alpha_s}{\pi}
\int_{Q_0}^Q\frac{dk_t}{k_t}C_F
\left(Y-i\pi\right)}
\mathcal{A}_0.
\end{equation}
It is easy to see that the Coulomb phase terms in the amplitude and its
conjugate cancel, having no physical effect.

\subsection{Gaps in \boldmath$2\to2$ scattering}

In $2\to2$ scattering, one can make an exactly analogous calculation,
with the one-loop result nesting and exponentiating to give the
all-order result.  The only difference is that for a hard process
involving more than three partons there can be more than one colour
structure, so the amplitude becomes a vector in colour space and the
loop correction (the $C_F(Y-i\pi)$ in the $e^+e^-$ case) becomes a
matrix,
\begin{equation}
\sigma=
\mathcal{A}_0^\dagger
e^{-\frac{2\alpha_s}{\pi}
\int_{Q_0}^Q\frac{dk_t}{k_t}\Gamma^\dagger}
S\,
e^{-\frac{2\alpha_s}{\pi}
\int_{Q_0}^Q\frac{dk_t}{k_t}\Gamma}
\mathcal{A}_0,
\end{equation}
where $S$ is the metric of the colour space.  The simplest case is quark
scattering, in which the colour space is 2 dimensional, and the
anomalous dimension matrix $\Gamma$ is given by\footnote{I am grateful
to Lev Lipatov for pointing out that this matrix was first calculated in
Ref.~\cite{Lipatov:1988ce}.}
\begin{equation}
\Gamma = \left(
\begin{array}{cc}
0&\frac{N_c^2-1}{4N_c^2}i\pi \\
i\pi & \frac{N_c}2Y-\frac1{N_c}i\pi
\end{array}\right).
\end{equation}
The important point is that $\Gamma$ and $\Gamma^\dagger$ do not
commute, so the Coulomb phase terms do not cancel.  Instead, they are
responsible for important physical effects, giving rise to the
`BFKL'-type logarithms in the limit of large $Y$\cite{Forshaw:2005sx}.

\subsection{Emission outside the gap}

The main point of Ref.~\cite{Forshaw:2006fk} was to check whether
emission outside the gap really cancels to all orders, as is observed in
the lowest order case, and as is assumed in the structure of the
all-order calculation.  To do this, we explicitly calculated the cross
section for one (real or virtual) gluon outside the gap, summed over any
number of soft virtual gluons integrated inside the gap and any number
of Coulomb gluons.  The result is simply the sum of the all-order
corrected virtual and real terms, integrated over the out-of-gap phase
space,
\begin{equation}
\sigma_1 = -\frac{2\alpha_s}{\pi}
\int_{Q_0}^Q\frac{dk_t}{k_t}
\int_{out}\frac{dy\,d\phi}{2\pi}
\left(\Omega_V+\Omega_R\right).
\end{equation}
$\Omega_V$ corresponds to one virtual emission outside the gap and its
all-order evolution.  It has a very similar structure to the
conventional gap cross section,
\begin{equation}
\Omega_V = \mathcal{A}_0^\dagger
e^{-\frac{2\alpha_s}{\pi}\int_{Q_0}^Q\frac{dk_t'}{k_t'}\Gamma^\dagger}
S_V
e^{-\frac{2\alpha_s}{\pi}\int_{Q_0}^{k_t}\frac{dk_t'}{k_t'}\Gamma}
\gamma\;
e^{-\frac{2\alpha_s}{\pi}\int_{k_t}^Q\frac{dk_t'}{k_t'}\Gamma}
\mathcal{A}_0
+c.c.,
\end{equation}
where $\gamma$ describes the virtual emission (roughly speaking it is
the differential of $\Gamma$) and I have just renamed $S$ to $S_V$ for a
reason that will be seen shortly.  The real part has a more complicated
structure, because it involves the evolution of a five-parton system at
scales below $k_t$ (the soft wide-angle gluon can be attached to the
real out-of-gap gluon, in addition to the original four partons).  The
five-parton colour structure has a different (higher) dimensionality
(four for the simplest case, $qq\to qqg$ for which the anomalous
dimension matrix, $\Lambda$, was calculated in
Ref.~\cite{Kyrieleis:2005dt}) so the real emission matrix element,
$D^\mu$, is a rectangular matrix acting on the colour space of the
four-parton process on the right and of the five-parton process on the
left.  The structure is then
\begin{equation}
\Omega_R = \mathcal{A}_0^\dagger
e^{-\frac{2\alpha_s}{\pi}\int_{k_t}^Q\frac{dk_t'}{k_t'}\Gamma^\dagger}
D_\mu^\dagger
e^{-\frac{2\alpha_s}{\pi}\int_{Q_0}^{k_t}\frac{dk_t'}{k_t'}\Lambda^\dagger}
S_R
e^{-\frac{2\alpha_s}{\pi}\int_{Q_0}^{k_t}\frac{dk_t'}{k_t'}\Lambda}
D^\mu\;
e^{-\frac{2\alpha_s}{\pi}\int_{k_t}^Q\frac{dk_t'}{k_t'}\Gamma}
\mathcal{A}_0,
\end{equation}
where $S_R$ is the metric of the five-parton colour space.

The out-of-gap gluon must be integrated everywhere outside the gap,
including right into the collinear regions, in which $\Omega_V$ and
$\Omega_R$ are separately divergent.  It is easy to check that in the
final-state collinear region, they indeed become
equal and opposite and the singularities cancel.  In the initial-state
collinear limit\footnote{This means the rapidity tending to infinity,
but at fixed $k_t$, so the emission never becomes truly collinear.}
however, one finds
\begin{equation}
\Omega_V+\Omega_R \stackrel{|y|\to\infty}\longrightarrow \mbox{const}.
\end{equation}
This means that in the pure eikonal theory the cross section is not
well-behaved, because the contribution from hard collinear
configurations becomes significant.  This non-cancellation can be traced
to the Coulomb phase terms in the evolution matrices, and ultimately to
the breakdown of \naive\ coherence discussed earlier.

Having made this discovery, it is easy to see how this behaviour leads
to the superleading logarithms we observed.  To leading approximation,
the effect of incorporating the correct splitting functions, energy
conservation, etc, in the collinear limit is to introduce an effective
cutoff on the rapidity range over which the eikonal result should be
integrated, $y_{max}\sim\ln\frac{Q}{k_t}$.  The nested integrals over
$k_t$ then have one additional log of $k_t$, leading to one additional
log of $Q/Q_0$,
\begin{equation}
\sigma_1\sim\sigma_0
\left(\frac{2\alpha_s}{\pi}\right)^4
\pi^2Y\,
\ln^5\frac{Q}{Q_0}
+\mathcal{O}\left(\alpha_s^n\ln^{n+1}\frac{Q}{Q_0}
\right).
\end{equation}

\section{Open Issues}

I end this talk by briefly mentioning some of the many open issues that
remain.

I stated that we do not need to consider soft gluons attached to the
hadron remnants.  A simple estimate shows that this has to be the case.
Since we are only interested in gluons with transverse momenta above
$Q_0$, even in the Glauber region, and we assume that $Q_0$ is large
relative to the hadronic scale, any such corrections should be
suppressed by powers of $Q_0$.  Nevertheless, since a number of
objections have been made in this direction, it would be worth working
through the first such correction, to shore up this argument.

Once we accept the breakdown of \naive\ coherence, the choice of
ordering variable becomes relevant.  Our calculation is based on
transverse momentum ordering and different ordering variables might give
different coefficients for the super-leading logarithms.  Further work,
for example by developing a full diagrammatic approach, is needed to be
sure that transverse momentum ordering leads to the correct physical
results.

Once we have found that one gluon outside the gap gives a tower of terms
enhanced by one additional logarithm, it is natural to speculate that
$n$ gluons outside the gap will give $n$ additional
logarithms\cite{Forshaw:2006fk}.  If this is right it would mean that at
each order, the leading term is actually $\as^n\log^{2n-3}$ and it would
be imperative to organize and sum these terms to all orders.  Performing
such a resummation is a daunting task, since, like an exact calculation
of non-global logarithms, it would depend on the full colour structure
of multi-parton ensembles.

I close by mentioning that I look forward to a critical appraisal of
this work.  The result came as such a surprise to us that we felt sure
it was wrong.  Two years of checking has not diminished this feeling.
However we have certainly ruled out simple error, since, in addition to
the two independent calculations we made, James Keates recently
succeeded in constructing an algorithm that generates all possible cut
diagrams order by order and evaluates them\cite{James}.  Within the same
strongly-ordered-in-$k_t$ approximation that we use, his calculation
reproduces ours up to fourth order, and confirms the coefficient of the
first super-leading logarithm.  Once issues of calculational speed have
been solved, he will be able to run at fifth order and beyond and check
our speculation about the r\^ole of multiple gluons outside the gap.

In the meantime we are trying to obtain a deeper understanding of our
findings, and I welcome any comments that help us in this direction.

\section*{Acknowledgements}

I am grateful to my collaborators Jeff Forshaw and Albrecht Kyrieleis,
and also to George Sterman, for fruitful discussions of these and
related points.

\begin{footnotesize}
\bibliographystyle{blois07} 
{\raggedright
\bibliography{blois07}

\providecommand{\etal}{et al.\xspace}
\providecommand{\href}[2]{#2}
\providecommand{\coll}{Coll.}
\catcode`\@=11
\def\@bibitem#1{%
\ifmc@bstsupport
  \mc@iftail{#1}%
    {;\newline\ignorespaces}%
    {\ifmc@first\else.\fi\orig@bibitem{#1}}
  \mc@firstfalse
\else
  \mc@iftail{#1}%
    {\ignorespaces}%
    {\orig@bibitem{#1}}%
\fi}%
\catcode`\@=12
\begin{mcbibliography}{10}

\bibitem{Forshaw:2006fk}
J.~R. Forshaw, A.~Kyrieleis, and M.~H. Seymour,
\newblock JHEP{} {\bf 08},~059~(2006).
\newblock \href{http://www.arXiv.org/abs/hep-ph/0604094}{{\tt
  hep-ph/0604094}}\relax
\relax
\bibitem{Dokshitzer:1985vs}
Y.~L. Dokshitzer and S.~I. Manaenkov.
\newblock LENINGRAD-85-1103\relax
\relax
\bibitem{Hakkinen:1996bb}
J.~Hakkinen and H.~Kharraziha,
\newblock Comput. Phys. Commun.{} {\bf 100},~311~(1997).
\newblock \href{http://www.arXiv.org/abs/hep-ph/9603229}{{\tt
  hep-ph/9603229}}\relax
\relax
\bibitem{Doria:1980ak}
R.~Doria, J.~Frenkel, and J.~C. Taylor,
\newblock Nucl. Phys.{} {\bf B168},~93~(1980)\relax
\relax
\bibitem{Collins:1985ue}
J.~C. Collins, D.~E. Soper, and G.~Sterman,
\newblock Nucl. Phys.{} {\bf B261},~104~(1985)\relax
\relax
\bibitem{Collins:1988ig}
J.~C. Collins, D.~E. Soper, and G.~Sterman,
\newblock Nucl. Phys.{} {\bf B308},~833~(1988)\relax
\relax
\bibitem{Dasgupta:2002bw}
M.~Dasgupta and G.~P. Salam,
\newblock JHEP{} {\bf 03},~017~(2002).
\newblock \href{http://www.arXiv.org/abs/hep-ph/0203009}{{\tt
  hep-ph/0203009}}\relax
\relax
\bibitem{Sotiropoulos:1993rd}
M.~G. Sotiropoulos and G.~Sterman,
\newblock Nucl. Phys.{} {\bf B419},~59~(1994).
\newblock \href{http://www.arXiv.org/abs/hep-ph/9310279}{{\tt
  hep-ph/9310279}}\relax
\relax
\bibitem{Oderda:1998en}
G.~Oderda and G.~Sterman,
\newblock Phys. Rev. Lett.{} {\bf 81},~3591~(1998).
\newblock \href{http://www.arXiv.org/abs/hep-ph/9806530}{{\tt
  hep-ph/9806530}}\relax
\relax
\bibitem{Berger:2001ns}
C.~F. Berger, T.~Kucs, and G.~Sterman,
\newblock Phys. Rev.{} {\bf D65},~094031~(2002).
\newblock \href{http://www.arXiv.org/abs/hep-ph/0110004}{{\tt
  hep-ph/0110004}}\relax
\relax
\bibitem{Lipatov:1988ce}
L.~N. Lipatov,
\newblock Nucl. Phys.{} {\bf B309},~379~(1988)\relax
\relax
\bibitem{Forshaw:2005sx}
J.~R. Forshaw, A.~Kyrieleis, and M.~H. Seymour,
\newblock JHEP{} {\bf 06},~034~(2005).
\newblock \href{http://www.arXiv.org/abs/hep-ph/0502086}{{\tt
  hep-ph/0502086}}\relax
\relax
\bibitem{Kyrieleis:2005dt}
A.~Kyrieleis and M.~H. Seymour,
\newblock JHEP{} {\bf 01},~085~(2006).
\newblock \href{http://www.arXiv.org/abs/hep-ph/0510089}{{\tt
  hep-ph/0510089}}\relax
\relax
\bibitem{James}
J.~Keates.
\newblock Work in progress\relax
\relax
\end{mcbibliography}
}
\end{footnotesize}
\end{document}